\documentclass[prl,twocolumn,letterpaper,superscriptaddress]{revtex4-1}
\usepackage[space]{grffile}
\usepackage{bm,graphicx,graphics,amsmath,amssymb,bm,epsfig,color}
\usepackage{euscript,tabularx}
\usepackage{longtable}
\usepackage{float}
\usepackage[draft]{todonotes}
\usepackage{tikz}

\def\<{\left\langle}
\def\>{\right\rangle}
\def\unit#1{{\hat{\bm#1}}}

\begin{document}

\title{Complete mapping of the spin-wave spectrum in vortex state
  nano-disk}

\author{B. Taurel}
\affiliation{INAC-SPINTEC, CEA/CNRS and Univ. Grenoble Alpes, 38000
  Grenoble, France}

\author{T. Valet} \affiliation{SPICE Johannes Gutenberg Universit\"at 55128
  Mainz, Germany}

\author{V. V. Naletov} 
\affiliation{INAC-SPINTEC, CEA/CNRS and Univ. Grenoble Alpes, 38000
  Grenoble, France}
\affiliation{Institute of Physics, Kazan Federal University, Kazan
    420008, Russian Federation}

\author{N. Vukadinovic} 
\affiliation{Dassault Aviation, 78 quai Marcel Dassault, 92552 St-Cloud, France}

\author{G. de Loubens} 
\affiliation{Service de Physique de l'\'Etat Condens\'e, CEA, CNRS,
  Universit\'e Paris-Saclay, CEA Saclay, 91191 Gif-sur-Yvette, France}

\author{O. Klein} 
\affiliation{INAC-SPINTEC, CEA/CNRS and Univ. Grenoble Alpes, 38000
  Grenoble, France}
\email[Corresponding author:]{ olivier.klein@cea.fr}

\date{\today}

\begin{abstract}
  We report a study on the complete spin-wave spectrum inside a vortex
  state nano-disk. Transformation of this spectrum is continuously
  monitored as the nano-disk becomes gradually magnetized by a
  perpendicular magnetic field and encouters a second order phase
  transition to the uniformly magnetized state. This reveals the
  bijective relationship that exists between the eigen-modes in the
  vortex state with the ones in the saturated state. It is found that
  the gyrotropic mode can be continuously viewed as a uniform phase
  precession, which uniquely softens (its frequency vanishes) at the
  saturation field to transform above into the Kittel mode.  By
  contrast the other spin-wave modes remain finite as a function of
  the applied field while their character is altered by level
  anti-crossing.
\end{abstract}

\maketitle

Magnetic equilibrium configurations adopting a singular topological
texture such as a vortex \cite{guslienko08}, bubble
\cite{komineas05,vukadinovic11}, or skyrmion
\cite{kim14,schwarze15} are currently attracting a lot of
attention as they allow engineering of the spin-wave (SW) spectrum
with potentially improved performances for spintronic devices
\cite{sampaio13,hamadeh14a}. An important feature is here the energy
density of states, which oversights inter-mode coupling. It has been
found that this coupling limits the performance of spin transfer
devices as it was shown in spin-orbit torque experiments
\cite{demidov11d,demidov12,hamadeh14b,collet16}. So far the most
effective topology that splits apart the bottom part of the energy
spectrum is the vortex ground state. It introduces a large gap between
the fundamental mode (the so-called gyrotropic mode)
\cite{guslienko02,park03} and the rest of the spectrum
\cite{novosad02,buess04}. The gyrotopic mode corresponds to a rotation
of the vortex core around its equilibrium position at the disk center
(the core being the small region where the magnetization points
out-of-plane). Because of near translational invariance for small aspect
ratio disks, it leads to a large renormalization of the associated
eigen-frequency, typically found to lie below 1~GHz
\cite{guslienko02}. Never the less, it remains coupled to the higher
part of the SW spectrum through an effective mass
\cite{guslienko10,guslienko15}. Despite numerous experimental
\cite{buess04,park05,aliev09,awad10,vogt11} and numerical
\cite{ivanov02,boust04,guslienko05a,bauer14} works on the dynamics of
the vortex state, a complete mapping of the SW spectrum above the
gyrotropic mode has not yet been proposed. Of particular concern is
the existence of hidden modes, whose odd symmetry shows no overlap
with a spatially uniform averaging \cite{naletov11}. In parallel
recent works on the sum rule invariance of the magnetic susceptibility
have underlined the continuity of the spectrum independently of the
magnetization texture \cite{thiaville12}. Still, this argument of
invariance has never been fully translated to a spectral point of view
due to the difficulty of establishing continuity between different
regions of the phase diagram.

\begin{figure}
  \includegraphics[width=8.5cm]{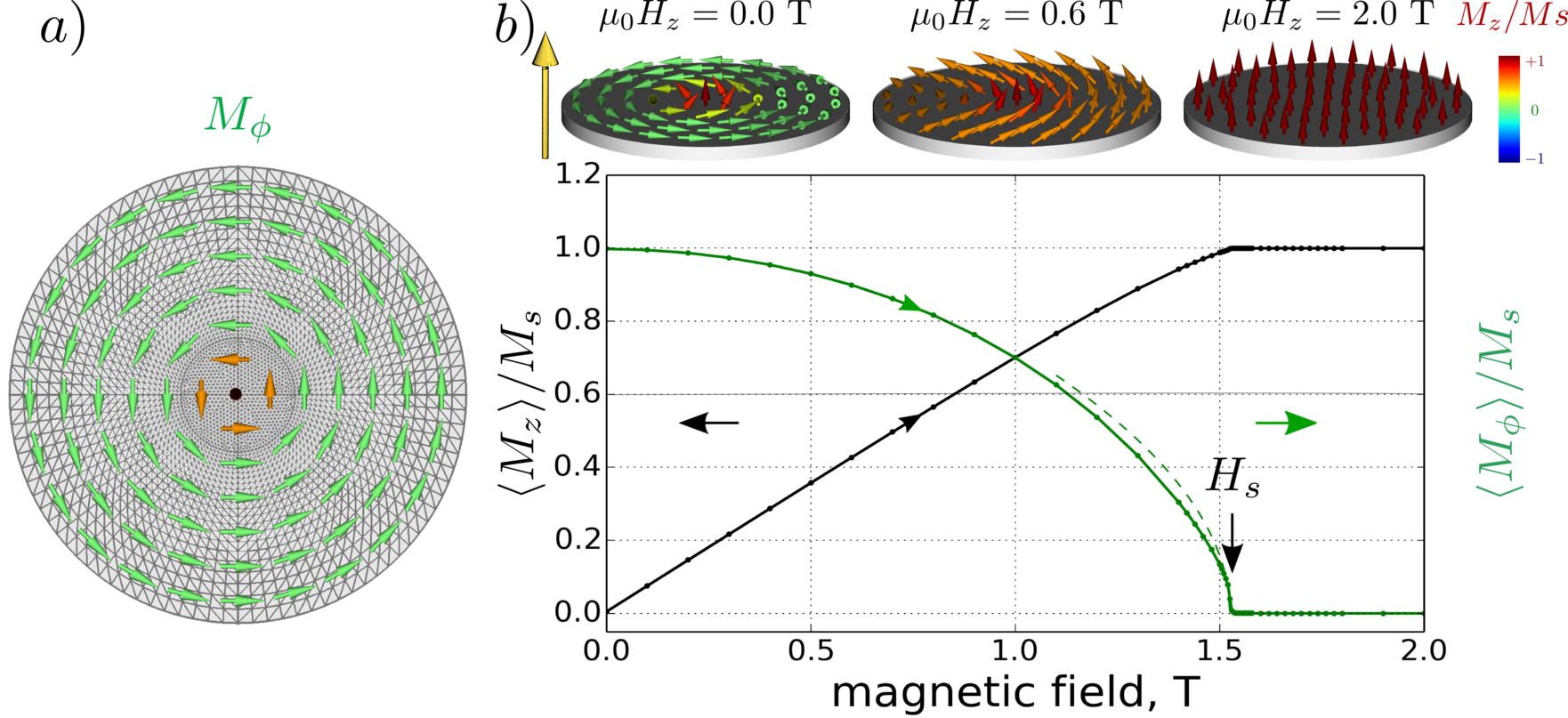}
  \caption{(Color online) a) Micromagnetic simulations performed on a
    vortex-state FeV nano-disk with $R=100$~nm radius, $t=10$~nm
    thickness, meshed adaptively around its center. b) Variation of
    its reduced longitudinal (black) and azimuthal (green)
    magnetization along an upward perpendicular magnetization cycle
    (the dashed line is the behavior of the order parameter for a
    critical exponent $= 0.5$). The above schematics show the
    equilibrium configurations simulated for three increasing values
    of the perpendicular field $H_z$. }
  \label{FIG1}
\end{figure}

In order to address this fundamental question, we propose to calculate
the SW spectra of a magnetic nano-disk along an upward perpendicular
magnetization cycle. The perpendicular configuration allows to monitor
the transformation of the SW spectrum as the magnetization texture
evolves continuously from the vortex state to the uniform state, by thus
avoiding any abrupt change of the equilibrium configuration
\cite{ivanov02,loubens09,castel12}. To fully account for the spectral
bijectivity, it is also important to calculate all possible
eigen-vectors, including the hidden modes aforementioned. This can
only be achieved by using an eigen-solver, which directly diagonalizes
the linearized Landau-Lifshitz equation.

In this work, we have used the SpinFlow3D micromagnetic simulation
package, a finite element simulation platform for spintronics, which
has been previously validated extensively for accurate computation of
eigen-modes in nano-objects \cite{naletov11,pigeau12}. All
micromagnetic simulations presented below are performed on the same
$R=100$~nm radius and 10~nm thickness disk, whose volume has been
meshed adaptively around the core singularity (see mesh in
Fig.~\ref{FIG1}a). The mesh-size is 1.6~nm in the central area of the
disk ($r<25$~nm) and is increased homothetically to reach 6~nm at the
periphery. The simulation is fully 3D thanks to three discretization
layers along the thickness to account for the potential texture in the
perpendicular direction (for a total of 8883 nodes). The magnetic
parameters used in the solver are the ones of FeV listed in
Table~\ref{tab:param} \cite{pigeau12}. A first numerical solver is
used to calculate the equilibrium state for different values of the
perpendicular magnetic field. It uses a Galerkin type finite element
implementation of the very efficient projection scheme introduced in
Ref.~\cite{e01}. Fig.~\ref{FIG1}b shows the normal magnetization cycle
produced by an upward perpendicular magnetic field swept between 0 and
2~T along $+z$. The initial state introduced in the simulation is a
vortex configuration, with the vortex core pointing towards $+z$. The
magnetization cycle shows an almost linear increase of the spatially
averaged normal component, $\langle M_z \rangle$, up to the saturation
field, which happens at $\mu_0H_s=1.525$~T. The growth of $M_z$ occurs
mainly through the canting of the peripheral spins which gradually tilt in
the perpendicular direction to form the so-called cone state
\cite{ivanov02} (see above schematic in Fig.~\ref{FIG1}b). By contrast,
the averaged azimuthal component, $\langle M_\phi \rangle$, which can
be viewed as the order parameter (see green curve in
Fig.~\ref{FIG1}b), vanishes continuously at $H_s$ indicating a second
order phase transition \cite{castel12}.

\begin{table}
  \caption{Parameters used in the simulation: magnetization,
    gyromagnetic ratio, exchange length, radius, thickness.}
  \begin{ruledtabular}
    \begin{tabular}{c c c c c}
      $\mu_0 M_s$ &  $\gamma$  & $\Lambda_\text{ex}$ &
      $R$  & $t$\\
      (T) & (rad$\cdot$s$^{-1}\cdot$T$^{-1}$) & (nm) & (nm) & (nm)\\
      \hline \\
      $1.7$ & $1.873 \times 10^{11}$ & 4.3 & 100 & 10
    \end{tabular}
  \end{ruledtabular}\label{tab:param}
\end{table}

\begin{figure}
  \includegraphics[width=8.5cm]{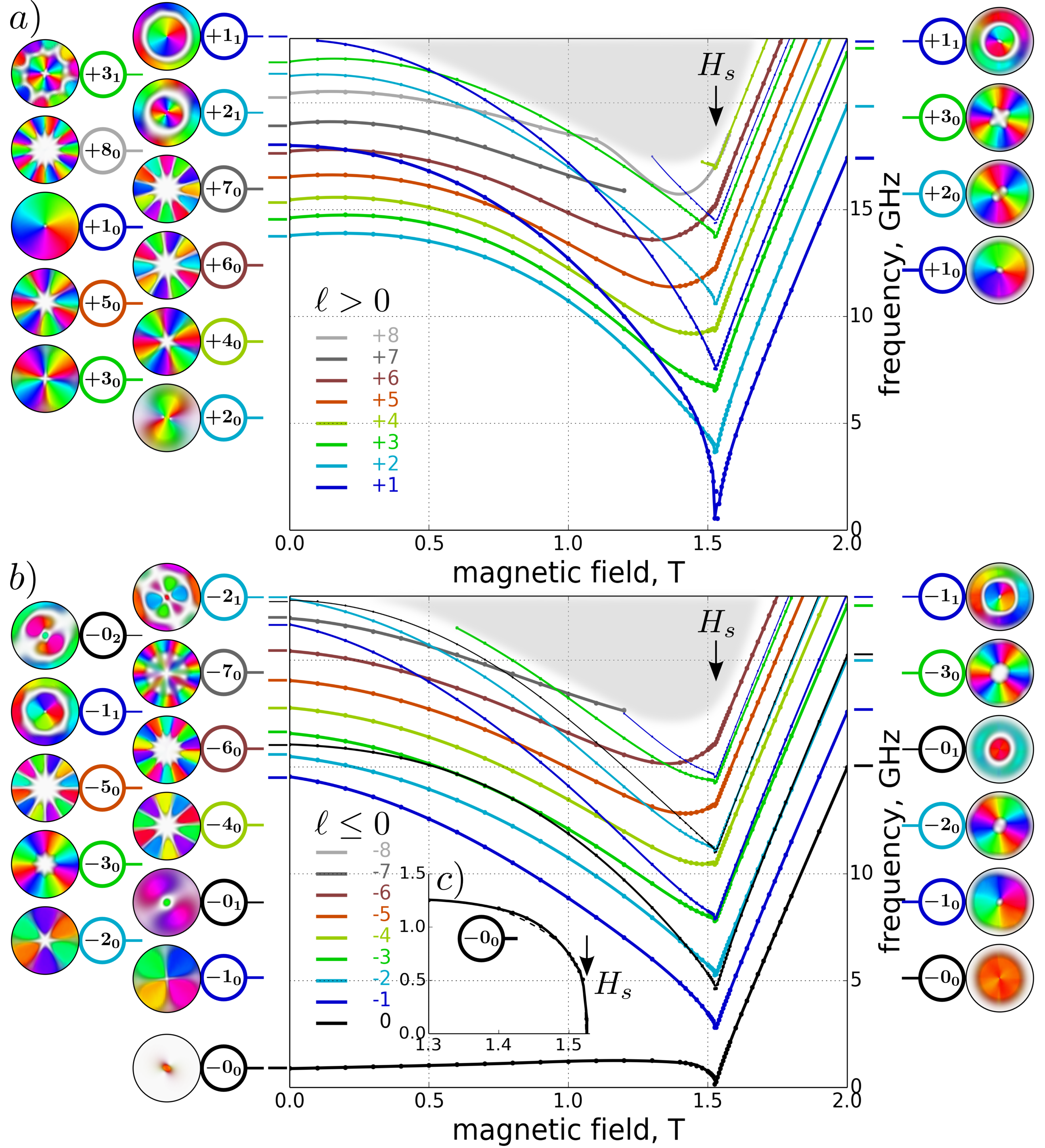}
  \caption{(Color online) Simulated spin-wave spectra of the nano-disk
    as a function of the perpendicular field $H_z$. Modes are labeled
    {$(\ell_m)$}, respectively the azimutal and radial indices. The
    two panels differentiate SW depending on their sense of gyration:
    a) $\ell > 0$ and b) $\ell \le 0$. Snapshot images of the
    eigen-vectors in the vortex ($\mu_0 H_z=0$~T) and saturated states
    (2~T) are shown on the left and right sides respectively
    (bi-variate color code, where amplitude/phase $=$
    luminance/hue). c) Zoom view of the softening of the gyrotropic
    mode at $H_s$ (the dashed line shows the dependence for a
    critical exponent $=0.3$).}
  \label{FIG2}
\end{figure}

Once the equilibrium magnetization is established at fixed values of
the applied field, a second numerical solver calculates the
eigen-states of the lossless linearized Landau-Lifshitz equation. It
solves the corresponding generalized eigen-value problem in the
vicinity of the pre-computed equilibrium state. It is solved with an
iterative Arnoldi method \cite{aquino09} using the ARPACK library
\cite{lehoucq98}. In this calculation the full complexity of the 3D
micromagnetic dynamics is preserved. The solver outputs in a few
minutes both the eigen-values by increasing order of energy and the
associated eigen-vectors. In this work, we have restricted the output
to the first 24 lowest energy modes. The results are displayed in
Fig.~\ref{FIG2}. We have separated the results in two panels depending
on the sense of gyration of the SW determined in reference to the core
polarity.

Three remarkable features are observed in Fig.~\ref{FIG2}. The first
one is the apparent one-to-one mapping between the eigen-modes in the
vortex state and the saturated state (the discontinuities observed at
the edge of the shaded area are just an artifact of the calculation
constraints, where only a finite number of modes are followed). The
second remarkable feature is the behavior of the lowest energy mode
shown by the continuous black line in Fig.~\ref{FIG2}b. This mode is
the gyrotropic mode below $H_s$ and the Kittel mode above. Its
frequency uniquely vanishes at $H_s$, while the eigen-modes above the
gyrotropic mode always resonate at finite frequency. Such behavior
seems generic to resonance modes in the presence of domain walls (see
chap. 8 of \cite{Gurevich1996}). The third one is that, contrary to
the saturated state, where the modes evolve in parallel as a function
of the applied magnetic field, in the vortex state the higher order
energy levels intercept each other. It suggests that a careful
analysis of the continuity of the mode character is required in the
open range $]0,H_s[$. In the following, we shall analyze in more
depths these three features.

We start with the identification of the SW spectrum. Several works
have studied the higher order modes in a vortex state nano-disks
\cite{ivanov02,buess04,guslienko05a}. In particular the spatially
averaged dynamic susceptibility tensor has been calculated
\cite{boust04}. In a recent review, we have performed such a
calculation for a downward perpendicular field sweep and a positive
field branch (see Fig.~4 in Ref.\cite{castel12}). It was found that
the in-plane component of the dynamic susceptibility varies in
strength along the cycle and almost disappears close to the saturation
field, leaving a blank window near $H_s$, which prevented the
establishment of the continuity between the two phases. While this
hurdle does not affect the output of SpinFlow3D shown in Fig.~\ref{FIG2}
and the continuity could be established by refining the field sweep
(small dots in Fig.~\ref{FIG2} indicate the different field values at
which a spectrum has been calculated), the character
of each SW mode in the vortex state remains to be assigned.

Finding the label of each eigen-state displayed in Fig.~\ref{FIG2}
requires to analyze the associated eigen-vectors.  Such an analysis
exists for the saturated state, where a labeling scheme has already
been proposed \cite{dillon60}. Above $H_s$, the eigen-vectors form the
complete set of Bessel functions \cite{klein08,naletov11} and each
eigen-state is fully described by two numbers {$(\ell_m)$},
respectively the azimutal and radial indices indicating the winding
numbers in these two directions. [Note that a third label indicating
the mode index along the thickness is not necessary here
\cite{zhou15}. For our 10~nm thin disk, one can safely consider that
all modes are uniform along the thickness. Higher order perpendicular
standing SW modes indeed occur outside the spectral range discussed
here.] Snapshot images of the precession profiles for the normally
magnetized disk are shown on the right side of Fig.~\ref{FIG2} using a
bi-variate coloring scheme: the hue codes the phase and the luminance
codes the amplitude. The winding numbers are inferred from the images
by counting respectively the number of times a color is
complementary/repeated in the radial/azimuthal directions.

\begin{figure}
  \includegraphics[width=8.5cm]{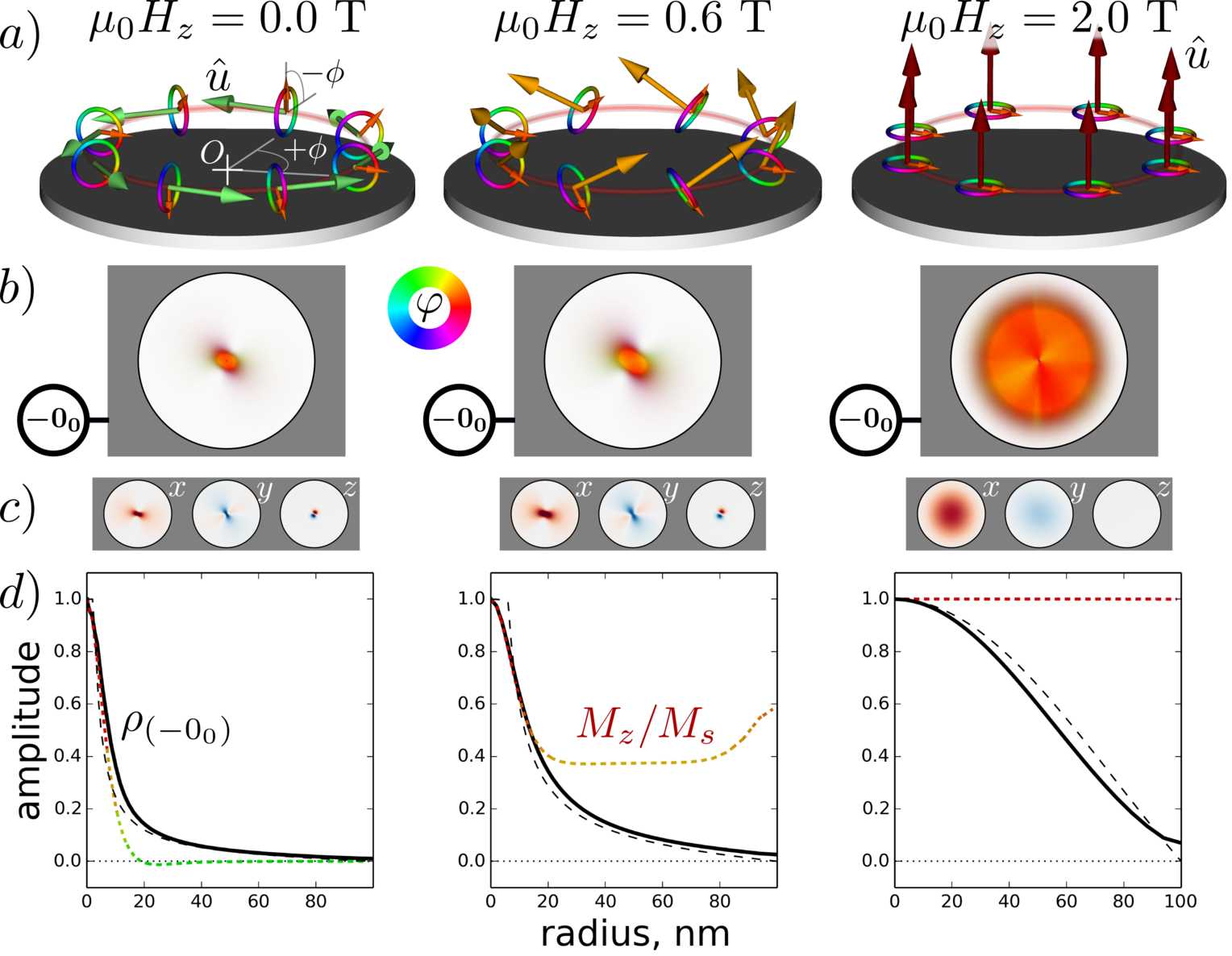}
  \caption{(Color online) a) Dynamical magnetization vector (short
    arrow) in the local frame of the magnetization texture (long
    arrow) for the mode with a uniform phase (phase $\varphi$ coded
    with the hue color wheel). b) Top view of the $(-0_0)$
    eigen-vector at $H_z=$0.0, 0.6, and 2.0~T using either the
    amplitude/phase representation or c) the cartesian directions
    (red-blue indicates region of opposite polarity). Radial profile
    of the normalized rms amplitude (continous line) compared with the
    analytical prediction (dashed line); the local value of the static
    magnetization $M_z$ is shown for comparison using a dotted line.}
  \label{FIG3}
\end{figure}

Below $H_s$ a local magnetic texture emerges and the unit vector
$\unit u$ pointing in the direction of the equilibrium magnetization
becomes dependent on the spatial coordinates (see
Fig.~\ref{FIG3}). The norm of the magnetization being a constant of
the motion, the possible eigen-vectors satisfy the local orthogonality
condition to $\unit u$. Possible directions for the dynamical
magnetization component are represented in Fig.~\ref{FIG3}a by a small
torus attached to the magnetization vector, whose phase reference
still needs to be properly defined. In the saturated state, the phase
reference is naturally a fixed cartesian direction. Hence the
transformation of the local frames between two positions separated by
the azimuthal angle $\phi$ occurs through a rotation of the reference
frame by an angle $+\phi$ around the disk center followed by a
rotation around $\unit u$ by an angle $-\phi$. Such transformation can
be generalized to any arbitrary texture conserving the axial
symmetry. For a vortex, the reference direction is shown by the small
arrow in Fig.~\ref{FIG3}a. At $\mu_0 H_z=0$~T the dynamical vector
for a mode with a uniform phase operates a full rotation in the
clockwise direction as one follows the curling magnetization
anti-clockwise along the periphery. We have put on the left side of
Fig.~\ref{FIG2} the zero field snapshot images associated with each
eigen-value output by the solver. The winding numbers of each image
can now be inferred from the color pattern. The eigen-values are
colored in Figs.~\ref{FIG2}a and \ref{FIG2}b according to the
indexation found (either a different color to dissociate SW having
different $|\ell|<9$ index or a different line thickness to dissociate
SW having different $m<4$ index). Extrapolating a straight line
between the different points underlines the existing relationship that
exists between the SW modes in the vortex and in the saturated states.

We concentrate now on the relationship that exists between the
gyrotropic mode and the Kittel mode (continuous black line at the
bottom of Fig.~\ref{FIG2}b). In our representation, both modes are
different manifestation of the same uniform phase albeit defined over
a background having different spatial texture. We have displayed in
Fig.~\ref{FIG3}b a top view of the $(-0_0)$ eigen-vector at three
values of the applied field: $\mu_0 H_z=$0, 0.6, and 2~T. We
stress that the phase of the mode $(-0_0)$, as represented in
Fig.~\ref{FIG3}b, is uniform throughout the volume, including in the
core region, where $\unit u$ points towards the normal direction. It
is interesting to note on the representations that the spatial average
of the in-plane component of the eigen-vector is always finite, while
by contrast, the averaged out-of-plane component vanishes. This
even/odd symmetry can be better observed by displaying the same
eigen-vectors along the cartesian directions (Fig.~\ref{FIG3}c). Thus
this mode only couples to a uniform rf excitation oriented
perpendicular to the core magnetization.  Another important feature to
notice in Fig.~\ref{FIG2}b is that, while the mode $(-0_0)$ remains
the lowest energy mode along the whole cycle, it is also the only mode
that softens at $H_s$. The softening of SW modes at the critical
fields between micromagnetic states has been investigated for the
cases of elliptical elements \cite{montecello07,montecello08} and
cylindrical nanodots with a large perpendicular anisotropy
\cite{vukadinovic11}. These critical fields correspond either to a
first-order or a second-order transition. In the former case, a
discontinuity in the magnetization curve occurs at the critical field
and the soft mode frequency goes to zero only on one side of the
transition (jump in the ${\omega}(H)$ curve at the critical field). In
the latter case, the magnetization curve is continuous at the critical
field and the soft mode frequency goes to zero continuously on both
sides of the transition. In both cases, the field dependence of the
soft mode angular frequency can be described by a power law in the
vicinity of the transition,
${\omega}\,{\propto}\,{\vert}H-H_{s}{\vert}^{\alpha}$, where
${\alpha}$ is the critical exponent. For elliptical elements
magnetized by a dc magnetic field applied along the minor axis, a
critical exponent ${\alpha=0.5}$ has already been reported
\cite{montecello08}. As noted above, in our case the transition
between the vortex state and the saturated state is of second
order. This can be evidenced by the behavior of the order parameter
which vanishes continuously at $H_s$ with a critical exponent $0.5$
(see green dashed line in Fig.~\ref{FIG1}a). In our case the soft mode
frequency vanishes like a power law with ${\alpha=0.3}$ (see black
dashed curve in Fig.~\ref{FIG2}c). This is the same exponent as the
one found for cylindrical nanodots with a perpendicular anisotropy in
the presence of a dc magnetic field along the symmetry axis, for the
soft mode existing at the transition between the bubble state and the
saturated state \cite{vukadinovic11}.

Since the phase pattern of the $(-0_0)$ mode is a conserved quantity
throughout the magnetization cycle, the only quantity that is field
dependent is the spatial distribution of the amplitude.  It is
interesting to note that top views of the precession profiles below
$H_s$ have a bow-tie shape. This feature is a direct consequence of
having a finite in-plane component, $M_\phi$. Indeed in
Fig.~\ref{FIG3}a, one can see that spins in orthogonal azimuthal
direction are orthogonal to each other. Since for small aspect ratio,
pointing out-of-plane is energetically defavorable compared to the
in-plane direction due to depolarization effects, the precession is
elliptical with the small axis in the normal direction. This
ellipticity is responsible for the observed amplitude modulation in
the azimuthal direction. We recall that the displayed images are
snapshots: time-wise the bow-tie rotates around the disk center at the
gyrotropic frequency. To account for this time dependence of the
amplitude, we have displayed in Fig.~\ref{FIG3}d the root mean square
(rms) amplitude averaged over one period. At zero field, one can
notice that the precession profile $\rho_{0_0}$ (black) extends well
outside the core region (shown by the dotted profile). An analytical
expression for the precession profile as been derived
\cite{guslienko08a} by Guslienko \textit{et al.} using the two vortex
ansatz $m_{0_0} \propto \left ( 1/\max(r_c,\rho) - \rho \right )$,
where $r_c$ is the core radius and all quantitites are expressed in
reduced units of $R$. It is displayed by a dashed line in
Fig.~\ref{FIG3}d; the predicted behavior agrees well with the
simulation. Its influence zone increases with the perpendicular
field. It reaches a maximum in the saturated state, where it
adopts a Bessel-function shape ($J_0$ dashed line) \cite{kakazei04}
with a node at the periphery because of dipolar pinning
\cite{guslienko05}.

\begin{figure}
  \includegraphics[width=8.5cm]{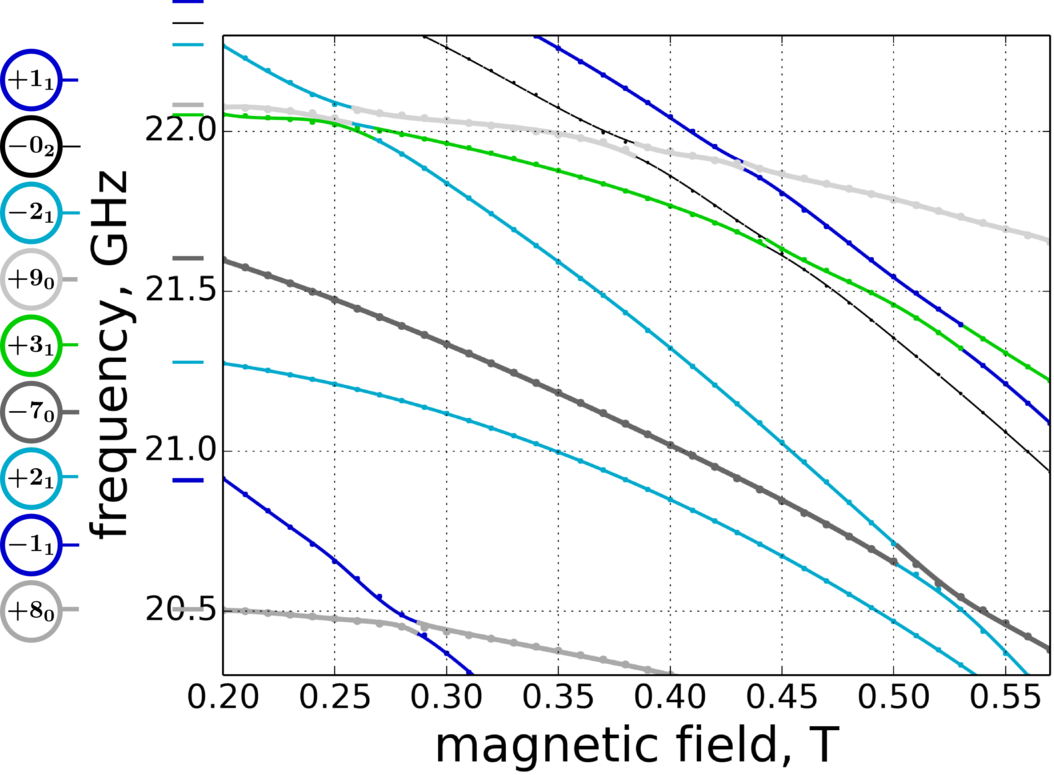}
  \caption{(Color online) Zoom view of the level anti-crossing between
    SW modes of different character in the vortex state.}
  \label{FIG4}
\end{figure}

Although the character of the fundamental mode remains unchanged, this
is not the case for the higher order SW modes, whose ranking changes
along the magnetization cycle due to crossing of energy levels. As
mentioned above, these crossings are responsible for the apparent
discontinuities in the upper part of the spectrum. We can note that
modes that bear the same $\ell$-index but different $m$-index never
cross each other. The opposite is not true, in particular between
pairs of opposite $\ell$-index.  This is best seen by putting on the
same plot the results shown in Fig.~\ref{FIG2}a and b.
Fig.~\ref{FIG4} is a zoom view of these combined spectra between 0.2
and 0.57~T. The zoom allows us to observe that the levels anti-cross
due to inter-mode coupling. It implies that the SW character can
hybridize in the vicinity of the near-degeneracy points and the SW
character can be modified by performing adiabatic minor cycles around
them. It implies also that at these anti-crossing the $\ell$ and
$m$-indices are not anymore good quantum numbers. Furthermore the
strength of this coupling depends obviously on the character of the
pair considered and the analysis of this underlying selection rule
remains to be done.

In summary, calculation of the eigen-states of a perpendicularly
magnetized disk has allowed us to establish the spectral relationship
that exists between eigen-modes in the vortex state and in the
uniformly magnetized state. This provides a complete mapping of the SW
spectrum above the gyrotropic mode, a particularly relevant result for
the understanding of the high energy regime of vortex-state
nano-objects, whose dynamics is governed by second order interaction with
these new possible modes.

\begin{acknowledgments}
  V. V. N. acknowledges support from the program CMIRA'Pro of the
  region Rh\^one-Alpes and from the Competitive Growth of KFU.
\end{acknowledgments}

%

\end{document}